\documentclass[a4paper,11pt]{article}
\pdfoutput=1 

\usepackage{jheppub} 

\usepackage[sort&compress]{natbib}
\usepackage[utf8]{inputenc}
\usepackage{graphicx}
\usepackage{subfig}
\usepackage{amsmath}
\usepackage{amssymb}
\usepackage{bm}
\usepackage{slashed}
\usepackage{paralist,array}
\usepackage{units}
\usepackage[dvipsnames]{xcolor}

\newcommand{\myparagraph}[1]{\bigskip \noindent\textbf{#1}}

\title{Connection between diphoton and triboson channels in new physics searches}

\date{}

\author[a]{Anastasia~Sokolenko,}
\author[b]{Kyrylo~Bondarenko,}
\author[b]{Alexey~Boyarsky}
\author[c]{and Lesya~Shchutska}

\affiliation[a]{Department of Physics, University of Oslo,Box 1048, NO-0371 Oslo, Norway}
\affiliation[b]{Intituut-Lorentz, Leiden University, Niels Bohrweg 2, 2333 CA Leiden, The Netherlands}
\affiliation[c]{Institute for Particle Physics and Astrophysics, Eidgen\"{o}ssische Technische Hochschule Z\"{u}rich, Otto-Stern-Weg 5, Z\"{u}rich, Switzerland}

\emailAdd{anastasia.sokolenko@fys.uio.no}
\emailAdd{bondarenko@lorentz.leidenuniv.nl}
\emailAdd{boyarsky@lorentz.leidenuniv.nl}
\emailAdd{Lesya.Shchutska@cern.ch}

\abstract{Diphoton channel provides a  clean signature in searches for new physics. In this paper, we discuss a connection between the diphoton channel ($\gamma\gamma$) and triboson channels ($Z\gamma\gamma$, $ZZ\gamma$, $WW\gamma$) imposed by the \mbox{$SU(2)_{L}\times U(1)_{Y}$} symmetry of the  Standard Model (SM) in certain classes of models. To illustrate this idea we choose a simple model that has all these channels. In this model, the same physics can give rise to $\gamma+$MET instead of $\gamma\gamma$ and 2 bosons plus missing energy instead of 3-boson channels. We analyze existing constraints and previous searches and show that channels $WW\gamma$ and especially $Z\gamma+$MET have a potential to discover new physics at the LHC.}

\begin{document}

\maketitle

\section{Introduction}

A diphoton signal is a good signature in the searches for new physics at the LHC~\cite{Khachatryan:2016yec,Aaboud:2017yyg} and  possible future colliders, for example, the ILC~\cite{Fujii:2015jha} or the FCC~\cite{fcc-design}. The diphoton channel was one of the first in the Higgs boson discovery~\cite{Aad:2012tfa,Chatrchyan:2012xdj}.
More recently, the unconfirmed 750~GeV resonance also appeared in the diphoton channel~\cite{ATLAS-diphoton,CMS:2015dxe,ATLAS-diphoton2016, ATLAS:2016eeo, CMS:2016crm}.

In this paper, we discuss the connection between the diphoton channel ($\gamma\gamma$)
and the three-boson channels ($Z\gamma\gamma$, $ZZ\gamma$, $WW\gamma$) that is imposed by the \mbox{$SU(2)_{L}\times U(1)_{Y}$} symmetry of the Standard Model (SM) for a certain class of models. The three-boson channels are interesting from experimental point of view because of low background and high detection efficiency~\cite{Binoth:2008kt,Bozzi:2009ig,Bozzi:2011en,Nhung:2013jta,Hong:2016aek}.
To illustrate this idea, we consider the specific axion-like particle model~\cite{Aparicio:2016iwr}.  Similar models were discussed in the context of the 750~GeV resonance that would, in this case, be explained by misidentification of a pair of photons created by a relativistic axion with a single photon due to the finite granularity of the detector~\cite{Knapen:2015dap,Bi:2015lcf,Chang:2015sdy,Agrawal:2015dbf,Chala:2015cev,Aparicio:2016iwr,Chen:2016sck}.

The paper is organized as follows: 
in Section~\ref{sec:themodel} we introduce a simple phenomenological model with a heavy scalar $s$ and light pseudo-Goldstone boson $a$ that can produce the corresponding signal. In  Section~\ref{sec:z_decay} we calculate constraints on the model coming from $Z$ boson decays. We discuss the 3-boson and 2-bosons-plus-missing-energy experimental signatures in Section~\ref{sec:discussion}, and conclude in  Section~\ref{sec:conclusion}.

\section{The model}
\label{sec:themodel}

Consider a simple extension of the SM with two new scalar particles, one of which is very light. This model naturally comes from the spontaneously symmetry breaking of a global $U(1)$ Peccei-Quinn symmetry~\cite{Peccei:1977hh} of a complex field $\phi$
\begin{equation}
 \phi = \frac{f + s}{\sqrt{2}} e^{ia/f},
\end{equation}
where $f$ is a vacuum expectation value of the $\phi$ field, $s$ and $a$ are real scalar fields. After the symmetry breaking one expecte the massive particle $s$ and the massless particle $a$ (the Goldstone boson). If the Peccei-Quinn symmetry is slightly broken, the field $a$ becomes massive, but in general much lighter that the heavy scalar particle $s$. The massive particle $a$ is called the \emph{axion}.
The interaction part of the Lagrangian is
\begin{equation}
 \mathcal{L}_{int} = -\frac{c_1}{2 f} a W_{\mu \nu}^i \widetilde{W}^{\mu \nu, i} 
 -\frac{c_2}{2 f} a B_{\mu \nu} \widetilde{B}^{\mu \nu} + s \frac{(\partial_{\mu} a)^2}{f} + \mathcal{L}_s,
 \label{eq:lagrangian}
\end{equation}
where $c_1$ and $c_2$ are dimensionless coupling constants, $B_{\mu \nu}$ and $W^i_{\mu \nu}$ are the strength tensors of the SM $U(1)_{Y}$ and $SU(2)_L$ gauge fields correspondingly. $\widetilde{W}^{\mu \nu, i}$ and $\widetilde{B}^{\mu \nu}$ are tensors dual to the strength tensors:
\begin{equation}
 \widetilde{F}_{\mu \nu} = \frac{1}{2} \varepsilon_{\mu\nu\sigma\rho} F^{\sigma\rho}.
\end{equation}
The Lagrangian $\mathcal{L}_s$ describes the   effective interaction of the $s$ particle with the SM. The interaction term between $s$ and $a$ comes from the kinetic term of the $\phi$ field, therefore it does not have additional coupling constant.

In terms of the physical fields, the structure of the gauge interaction terms is the following
\begin{equation}
  \mathcal{L}_{\text{gauge}} = \boxed{a \gamma \gamma} + \boxed{a Z Z} 
 + \boxed{a \gamma Z} + \boxed{a W^+ W^-} +  \boxed{a W^+ W^- \gamma} 
 + \boxed{a W^+ W^- Z},
 \label{eq:lagrangian_phys}
\end{equation}
where the part with 3 bosons is given by
\begin{align}
 \mathcal{L}_{aVV} = &-\frac{1}{4f} \epsilon^{\alpha \beta \gamma \delta} \Big[ (c_1 \sin^2\theta_W + c_2 \cos^2\theta_W) \, a F_{\alpha \beta} 
 F_{\gamma \delta} + (c_1 \cos^2\theta_W + c_2 \sin^2\theta_W) \, a Z_{\alpha \beta} Z_{\gamma \delta} \nonumber\\
 &- 2 \sin\theta_W \cos\theta_W (c_1 - c_2) \, a F_{\alpha \beta} Z_{\gamma \delta} + 2 c_1 a W_{\alpha\beta}^+ W_{\gamma\delta}^-
 \Big].
\end{align}

\begin{figure}[h]
    \centering
    \subfloat[]{\includegraphics[width=0.3\textwidth]{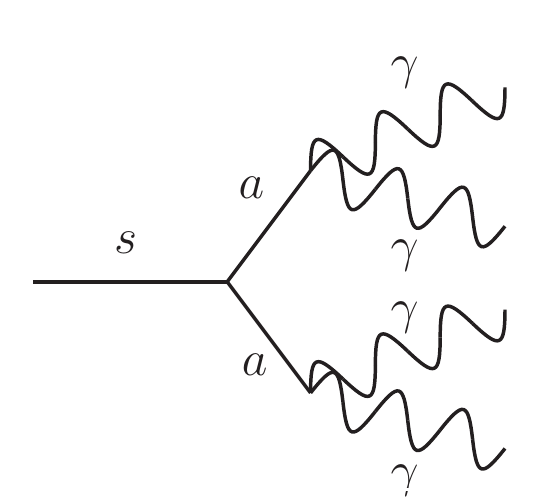}\label{fig:sdecay}}
    ~
    \subfloat[]{\includegraphics[width=0.3\textwidth]{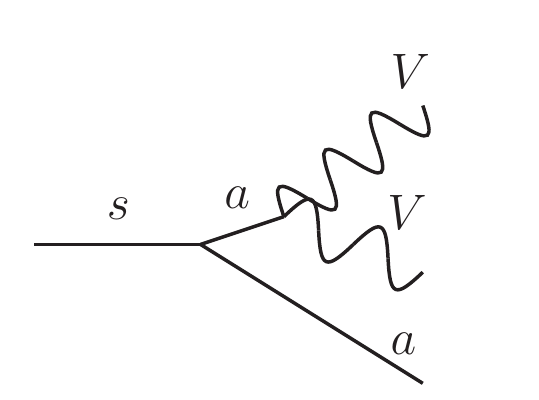}\label{fig:s2aVV}}
    \caption{(a): decay of the heavy scalar $s$ into 2 axions with subsequent decay into 2 photons. (b): decay of the heavy scalar into axion and 2 vector bosons.}
\end{figure}

In this model photon misidentification is possible for the $s$ decay shown in FIG. \ref{fig:sdecay}. The energy of the axion is at least $M_s/2$ so for low axion mass $m_a$ the misidentification of two photos as one happens if 
\begin{equation}
 \Delta \theta > \frac{12 m_a}{M_s},
\end{equation}
where $\Delta\theta$ is a granularity of the calorimeter, see formula~(\ref{eq:deltatheta}). In this case, this channel looks like a diphoton decay.

The gauge invariance requires existence of decays \mbox{$s\to a ZZ$}, \mbox{$s\to a Z\gamma$}, \mbox{$s\to a WW$}, that are connected to \mbox{$s\to aa\to 4\gamma$} decay. From the experimental point of view, these channels look like decays into 3 bosons: $\gamma ZZ$, $\gamma\gamma Z$ and $\gamma W W$. Although the 3 boson channels should have smaller branching ratio than $s\to aa$ decay, it is possible that they are more experimentally favourable. We will discuss such scenario below.

\subsection{Decays of the heavy scalar}
\label{sec:decays}

The main decay channel of the heavy scalar in the  model~\eqref{eq:lagrangian} is $s\rightarrow a a$. The decay width for this channel is
\begin{equation}
 \Gamma_{s \rightarrow a a} = \frac{1}{32 \pi} \frac{M_s^3}{f^2}.
 \label{eq:saa}
\end{equation}

\begin{figure}[h]
   \centering
   \includegraphics[width=0.5\textwidth]{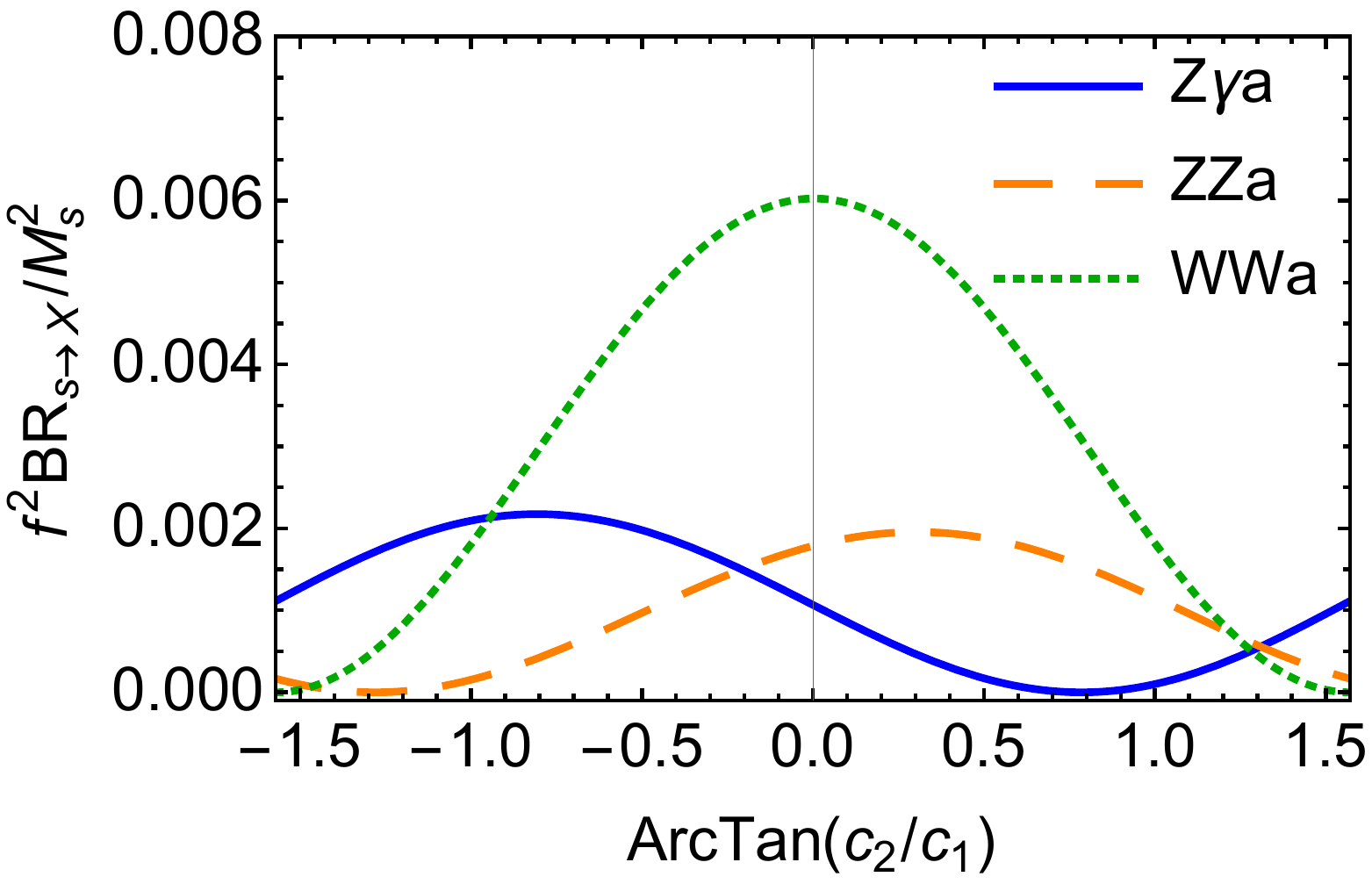}
   \caption{Branching ratios of the 3 body channels for different ratios between $c_1$ and $c_2$ coupling constants: continuous line is \mbox{$s\rightarrow Z \gamma a$}, dashed line is \mbox{$s\rightarrow ZZa$} and dotted line is \mbox{$s\rightarrow W W a$} channel. To make this plot we use the constraint \mbox{$c_1^2 + c_2^2 = 1$}.}
   \label{fig:3bodybr}
\end{figure}

From the Lagrangian~(\ref{eq:lagrangian_phys}) we expect the additional 3-boson decay channels: decay of $s$ into \mbox{$Z \gamma a$}, \mbox{$ZZ a$} or \mbox{$WW a$} (see FIG. \ref{fig:s2aVV}).
The decay widths in the limit \mbox{$M_s \gg M_Z, M_W$} are
\begin{align}
 &\Gamma_{s \rightarrow Z \gamma a} = 
 (c_1-c_2)^2 \sin^2 \theta_W \cos^2 \theta_W\frac{M_s^2}{16\pi^2 f^2} 
 \Gamma_{s \rightarrow a a},
 \label{eq:sZga}\\
 &\Gamma_{s \rightarrow Z Z a} =
 (c_1 \cos^2 \theta_W + c_2 \sin^2 \theta_W)^2
 \frac{M_s^2}{32 \pi^2 f^2} 
 \Gamma_{s \rightarrow a a}, 
 \label{eq:sZZa}\\
 &\Gamma_{s \rightarrow W W a} = c_1^2 
 \frac{M_s^2}{16\pi^2 f^2} 
 \Gamma_{s \rightarrow a a}.
 \label{eq:sWWa}
\end{align}
The branching ratios for these channels depend on the ratio between the coupling constants $c_1$, $c_2$, see FIG.~\ref{fig:3bodybr}.
For generic values of $c_1,c_2$ all three channels have branching ratios of the same order of magnitude.

\begin{figure}[h]
   \centering
   \includegraphics[width=0.5\textwidth]{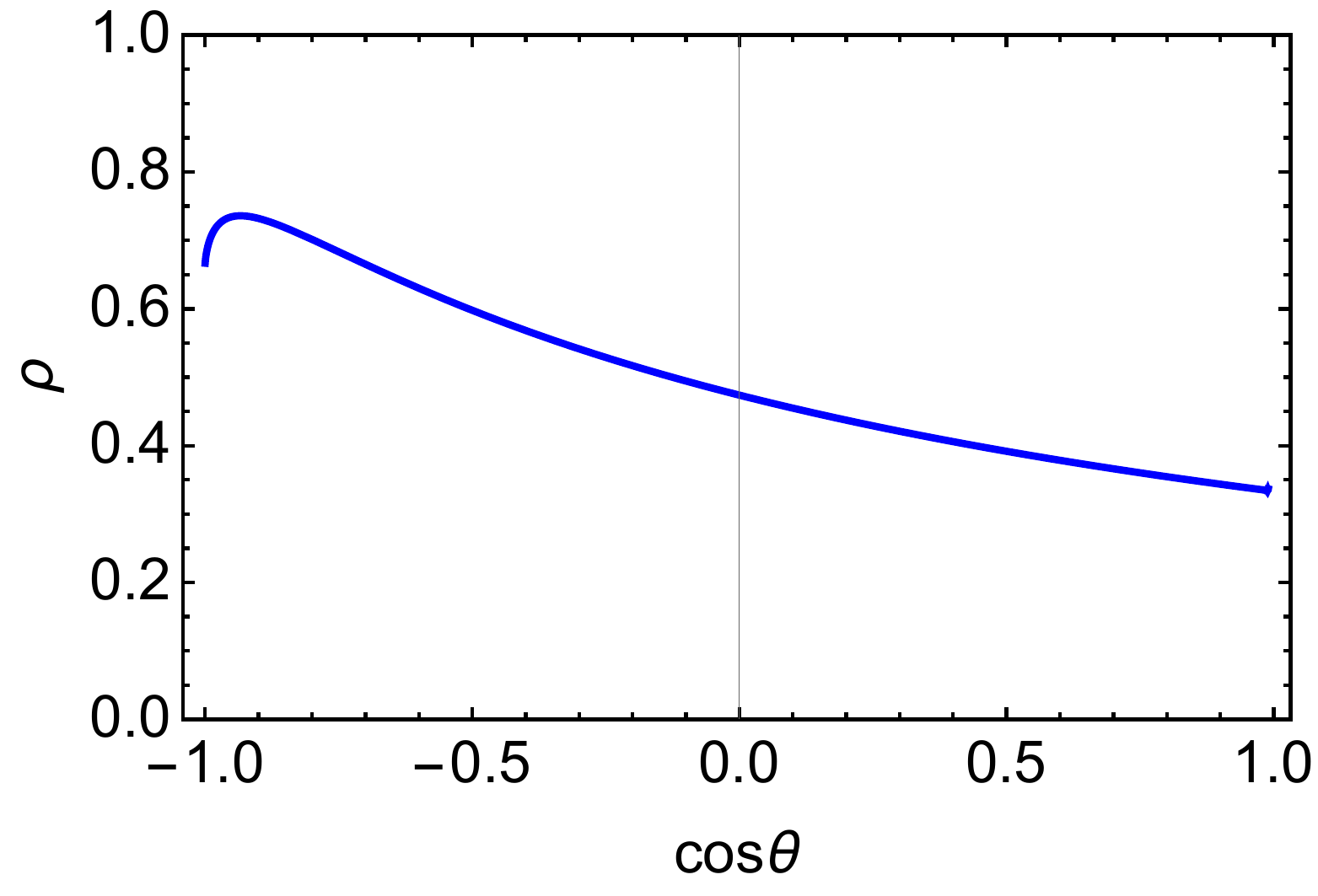}
   \caption{The angular distribution for the $s$ particle decay into three boson, where $\rho = \dfrac{1}{\Gamma} \dfrac{d\Gamma}{d\cos\theta}$ and $\theta$ is an angle between the vector bosons ($Z\gamma$, $ZZ$ or $WW$).}
   \label{fig:angle}
\end{figure}

All three channels have similar angular distributions for the vector bosons. These distributions are equal to each other in the limit \mbox{$M_s\gg M_Z, M_W$}. The angular distribution for this case is presented in FIG.~\ref{fig:angle}. We see that vector bosons prefer to fly in opposite directions. The average angle between them is 
\mbox{$\theta\approx 98^{\circ}$}.

\begin{figure}[h]
   \centering
   \includegraphics[width=0.5\textwidth]{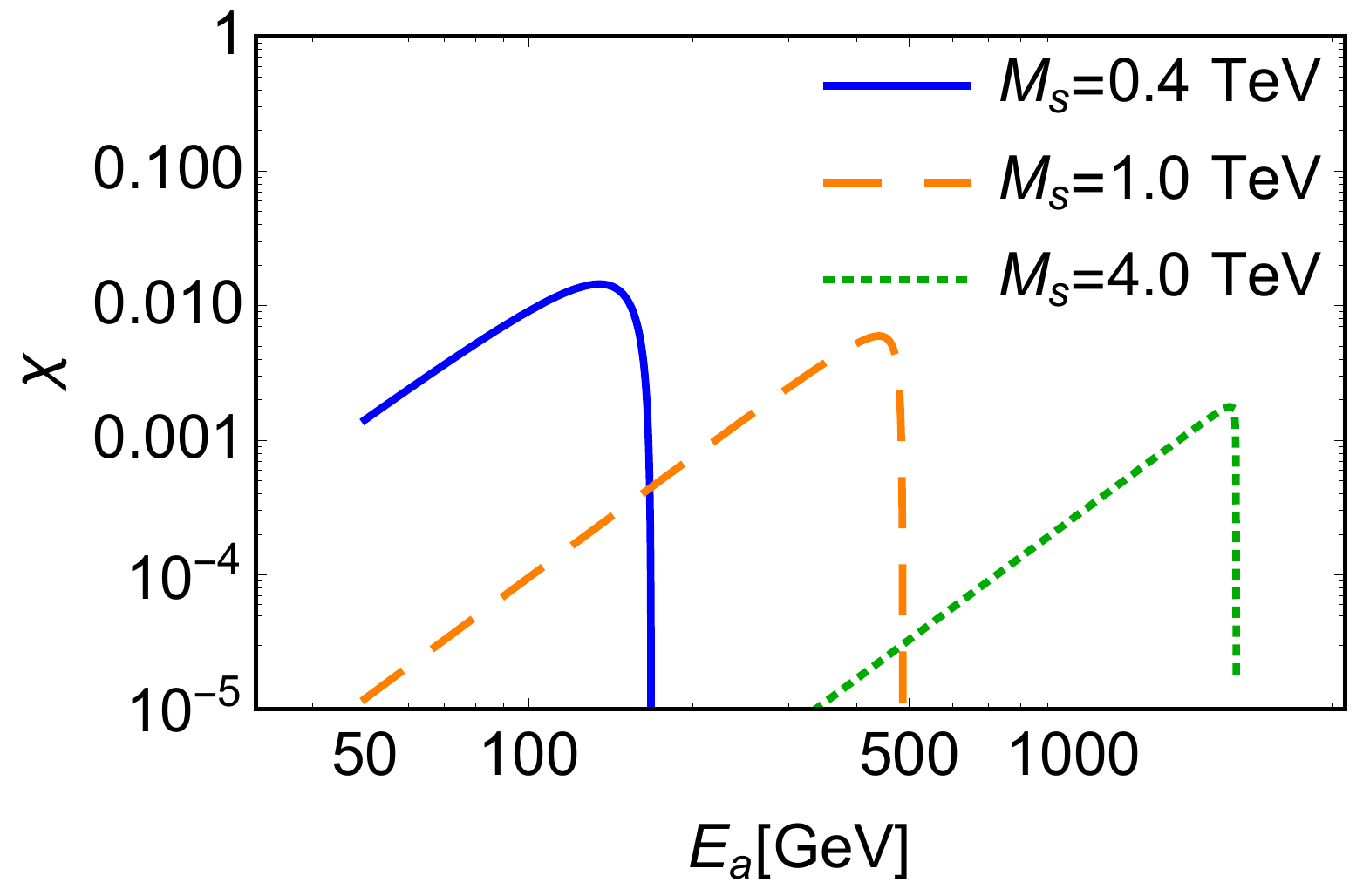}
   \caption{The energy distribution $\chi$ for the $s$ particle decay into $WWa$ by axion energy $E_a$, where $\chi=\dfrac{1}{\Gamma} \dfrac{d\Gamma}{dE_a}$.}
   \label{fig:energydistribution}
\end{figure}

FIG.~\ref{fig:energydistribution} shows the axion energy distribution $\dfrac{1}{\Gamma} \dfrac{d\Gamma}{dE_a}$ for the process $s \to WWa$ for 3 different masses of $s$ particle. At the low axion energy $E_a\ll M_s$ the distribution scales as
\begin{equation}
 \dfrac{d\Gamma}{dE_a} \propto E_a^3
\end{equation}
and the cut-off energy is $E_{a}^{\max} = \dfrac{M_s^2 - 4 M_W^2}{2 M_s}$.

\section{Existing constraints on the model}
\label{sec:z_decay}

\begin{figure}[h]
  \centering
  \includegraphics[width=0.3\textwidth]{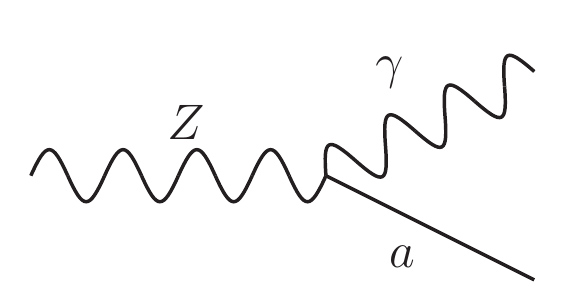}
  \caption{Decay of $Z$ boson into a photon and the axion. If a boosted photon pair is mis-identified as a single photon, this decay would look as $Z \to \gamma\gamma$ experimentally.}
  \label{z}
\end{figure}

The strongest constraints on the parameters come from the precision measurements of $Z$.
In our model a new decay channel of $Z$ boson appears (see FIG.~\ref{z}). The decay width is given by
\begin{equation}
 \Gamma_{Z \rightarrow a \gamma} =\frac{1}{96 \pi f^2} (c_1 - c_2)^2  \sin^2 2\theta_W M_Z^3,
 \label{eq:Zagamma}
\end{equation}
where we neglect the mass of the axion. 

After the axion decay, we have 3 photons with small opening angle $\Delta\theta_Z$ between two of them, produced from the axion. The energy of the axion is at least $M_Z/2$. Thus, using formula~(\ref{eq:deltatheta}) the constraint on the opening angle is
\begin{equation}
 \Delta\theta_Z \le \frac{12 m_a}{M_Z}.
 \label{eq:deltathetaZ}
\end{equation}

It is interesting to mention that $Z$ boson decay into two photons is forbidden by the Landau-Yang theorem, mentioned above. Nevertheless, the idea that $Z$ boson can produce 2 photon decay signature through the light pseudoscalar particle is not new. There is a SM decay $Z\rightarrow \pi^0 \gamma$ with expected branching ratio from $10^{-12}$ to $10^{-9}$~\cite{Deshpande:1990ej, Hikasa:1990xx, Manohar:1990hu, Pham:1990mf, West:1990ta, Young:1990pt, Schroder:1990nr, Chatterjee:1990eg, Maitra:1994pm, Micu:1995sc}. The decay of this type was searched before~\cite{Aaltonen:2013mfa}, but not at the LHC.

The measurement of the $Z$ boson decay into 2 photons was performed by the CDF collaboration~\cite{Aaltonen:2013mfa} providing an upper bound
\begin{equation}
 \text{BR}(Z\rightarrow \gamma\gamma) \le 1.5\cdot 10^{-5}.
 \label{eq:zbound}
\end{equation}
The  angular resolution of the CDF calorimeter is $\Delta \theta_{\text{CDF}} \ge 0.1$~\cite{Blair:1996kx}. It is lower than the maximal opening angle~(\ref{eq:deltathetaZ}) if $m_a \lesssim 750$~MeV, so the model~\eqref{eq:lagrangian} would produce a diphoton signature of $Z$ boson decay in this case.  The bound~(\ref{eq:zbound})  constraints  the model parameters to be
\begin{equation}
 \frac{|c_1 - c_2|}{f} \le 1.6\cdot 10^{-4}\text{ GeV}^{-1} .
 \label{eq:Zconstraint}
\end{equation}
Another independent constraint comes from the full decay width of the $Z$ boson. Value of total decay width of the $Z$ boson is measured as $\Gamma_Z^{\text{exp}} = 2.4952(23)\text{ GeV}$~\cite{Agashe:2014kda}. 
It is equal to the SM theoretical prediction $\Gamma_Z^{\text{SM}}=2.4960(18)\text{ GeV}$~\cite{Novikov:1999af,Schael:2013ita} within experimental uncertainties. We estimate $1\sigma$ deviation from the $Z$ decay width as
\begin{equation}
 \Delta\Gamma_Z = \sqrt{\Delta\Gamma_{Z,\text{exp}}^2 + \Delta\Gamma_{Z,\text{SM}}^2} = 2.9\text{ MeV}
\end{equation}
and require that decay width of new channel $Z\to a\gamma$ is within $2\sigma$ limit,
\begin{equation}
 \frac{|c_1 - c_2|}{f} \le 1.8\cdot 10^{-3} \text{ GeV}^{-1}.
\end{equation}
The last constraint is weaker than~\eqref{eq:Zconstraint}, but it does not depend on the detection of the axion as one photon.

\section{Results}
\label{sec:discussion}

\subsection{Sensitivity of the triboson vs. diphoton channels}
\label{sec:discussionA}

In this Section we consider triboson channels that arise from \mbox{$s\rightarrow a VV$} decays. The experimental signatures in these channels are: $Z\gamma\gamma$, $ZZ\gamma$ and $WW\gamma$, where the vector bosons are not collimated (c.f.\ FIG.~\ref{fig:angle}). 
We analyze the sensitivity to these channels, given current constraints on the diphoton searches.

We start with the decays containing $Z$ boson.
The final states of leptonically decaying $Z$ bosons have lower SM background as compared to the hadronic decays. The probability of the $Z$ boson decay into $e^+e^-$ or $\mu^+\mu^-$ is $P_{Z\rightarrow l^+ l^-} = 6.7\%$ (we do not take into account \mbox{$Z\rightarrow\tau^+\tau^-$} because it is reconstructed through hadronic $\tau$ decays with high SM background). Therefore for generic values of $c_1,c_2$ the channel $Z\gamma\gamma$ is more favorable to search  than $ZZ\gamma$. The $W$ boson cannot be fully reconstructed in the leptonic decay mode. 
Thus we conclude that $Z\gamma\gamma$ channel is the most sensitive among the three considered. 

The main background in the $Z\gamma\gamma$ channel comes from the non-resonant SM $Z\gamma$\ production, which has quite a low production cross section in the phase space of interest. Comparing the measured SM backgrounds in papers describing the searches in the $Z(l\bar{l})\gamma$ channel~\cite{Sirunyan:2017hsb} and in the diphoton channel~\cite{CMS:2016crm}, we see that $Z\gamma$ channel has an order of magnitude lower background than diphoton one. This background is even further suppressed by the requirement of an additional energetic photon in the event. Therefore, we expect that this channel is almost background-free.

From Eqs.~(\ref{eq:saa}) and (\ref{eq:sZga}) and from the constraint~(\ref{eq:Zconstraint}) we find the following limit on the branching ratio
\begin{equation}
 \text{BR}(s\rightarrow Z\gamma a) \le 1.5\cdot10^{-5}
 \left( \frac{M_s}{750\text{ GeV}} \right)^2.
\end{equation}
Consider that we expect 1 event in this channel. Then, taking into account the probability of $Z$ decay into charged leptons, we expect 
\begin{equation}
 N \ge 10^6 \left(\frac{750\text{ GeV}}{M_s} \right)^2
\end{equation}
events in the diphoton channel. This number cannot be covered up by any reasonable SM background, therefore $Z\gamma\gamma$ channel is less sensitive than the diphoton one.
The conclusion above is also valid for $WW\gamma$ and $ZZ\gamma$ channels if there is no degeneracy. 

In case of the degeneracy \mbox{$c_1 \approx c_2$}, the $Z\gamma\gamma$ channel is suppressed and \mbox{$\Gamma_{s\rightarrow WWa} = 2\Gamma_{s\rightarrow ZZa}$} (see expressions~(\ref{eq:sZZa}) and (\ref{eq:sWWa})). 
The number of events in diphoton channel $N_{\gamma\gamma}$ is connected to the number of events in $WW\gamma$ channel,
\begin{equation}
 \frac{N_{WW\gamma}}{N_{\gamma\gamma}} = 
 \frac{\Gamma_{s\rightarrow WWa}}{\Gamma_{s\rightarrow aa}} = \frac{M_s^2 c_1^2}{16 \pi^2 f^2}.
 \label{eq:gdivg}
\end{equation}

One can search for $WW\gamma$ signature in two final states where either only one W boson decays leptonically ($W\to e \nu$ or $W\to\mu\nu$), or both W bosons decay to leptons. In the first case, the main SM background comes from the $W\gamma$ production with two additional jets, where these two jets accidentally form a $W$ boson mass. The number of background events rapidly drops with the increase of the photon transverse momentum $E_T^\gamma$, and is equal to about 1 event for $E_T^\gamma > 300$ GeV. From the parton luminosity scaling for quark-annihilation processes between center-of-mass energies of 8 and 13 TeV, the corresponding number of background events should be about a factor of 2 larger for the same integrated luminosity, and factor 3 larger for the integrated luminosity delivered by the LHC in 2016. Such background rate would lead to an upper limit on a number of signal events in the range from 3 (for the zero background case) to 6 (for a number of background events equal to 3) for the mass $M_s > 1$ TeV. This converts into and the upper limit on the signal cross section of about 0.3-0.6 fb. In this estimate, the branching ratio correction of 0.3 is taken into account, and it is assumed that signal has 100\% reconstruction and identification efficiency. 

In the second case, when both $W$ bosons decay leptonically, the main SM backgrounds arise from $t\bar{t}\gamma$, $Z\gamma$, $WZ\gamma$ processes, 
and processes with a misidentified photon. The SM background becomes negligible for $E_T^\gamma > 300$ GeV, hence we can conduct the estimates in a zero background approximation. The branching ratio correction for this scenario would be 0.06, and this leads to factor 5 weaker constraints on the signal cross sections compared to the semileptonic $WW\gamma$ channel. 

Let us discuss the possibility to observe 3-boson channel before the diphoton one. This is possible if the number of the background events in the diphoton channel $N_{\gamma\gamma}^{\text{bg}}$ is much higher than the background in the 3-boson one $N_{WW\gamma}^{\text{bg}}$. The condition in the case of the Gaussian statistics reads as
\begin{equation}
    \frac{N_{WW\gamma}}{N_{\gamma\gamma}} > \left(\frac{N_{WW\gamma}^{\text{bg}}}{N_{\gamma\gamma}^{\text{bg}}}\right)^{1/2}
    \label{eq:conditiontosee3V}
\end{equation}

\begin{figure}[h]
   \centering
   \includegraphics[width=0.5\textwidth]{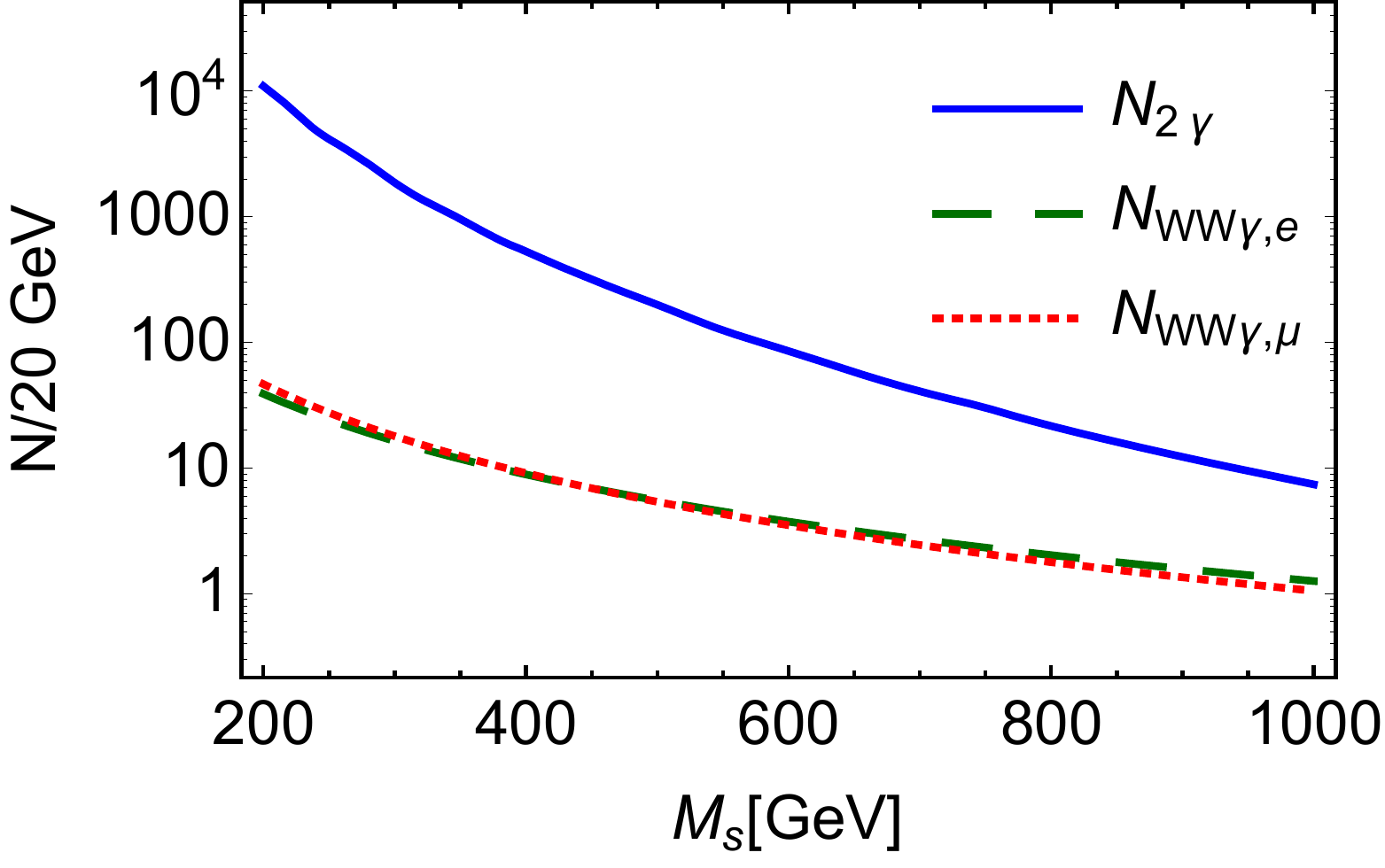}
   \caption{Estimation of the number of background events for the $\sqrt{s}= 13$ and integrated luminosity of $\mathcal{L} = 36.7\text{ fb}^{-1}$ for diphoton (blue solid line), $WW\gamma \to W e \nu \gamma$ (green dashed line) and $WW\gamma \to W \mu \nu \gamma$ (red dotted line) channels.}
   \label{fig:bkg}
\end{figure}

The data on diphoton background can be found in the paper~\cite{Aaboud:2017yyg} by the ATLAS Collaboration, where the bounds on the peak search of the diphoton signal are given at $\sqrt{s}= 13$~TeV with integrated luminosity of $\mathcal{L}_{0} = 36.7\text{ fb}^{-1}$. Experimental analysis of the $WW\gamma$ signature has been performed by the ATLAS Collaboration at the center-of-mass energy of $\sqrt{s} = 8$~TeV in the context of the measurement of the SM $WW\gamma$ production and search for anomalous quartic gauge couplings~\cite{Aaboud:2017tcq}. From this paper we can extract the background in the case of when only one W boson decays leptonically ($W\to e \nu$ or $W\to\mu\nu$).
Adopting these backgrounds for the same center-of-mass energy and the same binning we get estimation shown at the FIG.~\ref{fig:bkg}. The background ratio is the largest for a small mass of the mass of the heavy scalar.

The ratio in the left-hand-side of the formula~\eqref{eq:conditiontosee3V} depends on parameters of the model. In Appendix~\ref{sec:toymodel} we discuss the simple UV-completion with $N_{\chi}$ heavy fermions. The natural value of constants are $M_s\sim f$ and $c_1\sim \alpha_{w} N_{\chi}$, so the estimate of the ratio~\eqref{eq:gdivg} is $N_{WWa}/N_{\gamma\gamma} \sim \alpha_w^2 N_{\chi}^2/(16\pi^2)$. So we make a conclusion that to observe 3-boson channel before diphoton one if possible is that number of heavy fermions is large enough.

\subsection{Axion as missing energy}
\label{sec:MET}

In the discussion before we have made an assumption that an axion decays inside the detector. In this section we  consider the case, that an axion could leave the detector, i.e. the decay length $l = c\gamma\tau$ (where $\tau$ is an axion lifetime and $\gamma$ is a Lorentz factor) is greater than detector length $L$. The decay length is (see Appendix~\ref{sec:axion_decay_length})
\begin{equation}
 l \approx 5\text{ m} \left( \frac{100\text{ MeV}}{m_a}\right)^4
 \left( \frac{M_s}{1 \text{ TeV}}\right)
 \left( \frac{f \cdot 10^{-4} \text{ GeV}^{-1}}{c_1 \sin^2 \theta_W + c_2 \cos^2 \theta_W} \right)^2.
\end{equation}
There exists the allowed parameter region where $l\gg L$, for example if axion mass $m_a<100$~MeV.
In this case the probability of the axion decay inside the detector is
\begin{equation}
 P_{\text{axion decay}} = 1 - e^{-L/l} \approx \frac{L}{l} \ll 1.
 \label{eq:Padecay}
\end{equation}

Instead of comparing diphoton channel with 3-boson channels as in the previous section, we have to compare the channel $\gamma+$MET~\cite{Sirunyan:2017ewk,Aaboud:2017dor} with  $ZZ$+MET, $WW$+MET and $Z\gamma$+MET channels. The search of the $ZZ$+MET, $WW$+MET signatures was performed at the LHC for the SUSY models~\cite{ATLAS:2017uun,Sirunyan:2018ubx}. The advantage of our model compared to SUSY case is that invariant mass of the decay products should be fixed by the mass of the heavy scalar. However, this fact does not give a significant improvement in the analysis, because one cannot measure the parallel component of the momentum for the missing energy.

Let us consider the channel $Z\gamma$+MET, as the dedicated searches were not performed at the LHC before. Let us check if the new channel can show a signal before $\gamma$+MET~\cite{Sirunyan:2017ewk,Aaboud:2017dor}. Using Eqs.~\eqref{eq:saa}, \eqref{eq:sZga} and~\eqref{eq:Padecay} the ratio of the probabilities of the signature $Z\gamma$+MET to $\gamma$+MET is
\begin{equation}
 \frac{P_{Z\rightarrow l^+ l^-} \Gamma_{s\to Z\gamma a}}{2 P_{\text{axion decay}} \Gamma_{s\to aa}} 
 \approx 
 1.9\cdot 10^{-6} \left( \frac{100\text{ MeV}}{m_a}\right)^4
 \left( \frac{M_s}{1 \text{ TeV}}\right)^3
 \left( \frac{c_1  - c_2}{c_1 \sin^2 \theta_W + c_2 \cos^2 \theta_W} \right)^2.
\end{equation}
We see that for $m_a\ll 100$~MeV and/or 
\mbox{$|c_1 \sin^2 \theta_W + c_2 \cos^2 \theta_W|\ll|c_1 - c_2|$} (e.g. $c_1$ and $c_2$ have opposite sign)
this ratio can be greater than one.
As $Z\gamma$+MET is expected to have lower background than $\gamma$+MET, so the former channel can be more efficient for the search of new physics at the LHC.

\section{Conclusion}
\label{sec:conclusion}

In this paper, we discussed the triboson channels as a potential signature of new physics at the LHC and analyze the corresponding sensitivity. 
We show that if a new particle decays to four photons,  with two collimated photons being misidentified as one photon and hence leading to a pick in the observed diphoton events, the gauge invariance of the SM demands the existence of additional decay channels the type $Z\gamma\gamma$, $ZZ\gamma$ and $WW\gamma$.

To illustrate this idea we choose the simple model with a heavy scalar $s$ 
and a light pseudoscalar $a$. We calculate the particle decay widths 
in this model and analyze the kinematic properties of triboson decays, 
see FIGs.~\ref{fig:angle}, \ref{fig:energydistribution}.

We find that the effective coupling $Za\gamma$ in this model is strongly constrained by the $Z\to \gamma\gamma$ decay searches, therefore we make a specific choice of model parameters $c_1=c_2$ to avoid this constraint. 
In this case, one still has significant freedom in the choice of remaining parameters.

The main advantage of the triboson channels is the lower value of expected SM background in comparison to the diphoton channel. Combining this property with the number of diphoton background event we conclude that this channel can be helpful for the searches in the region of the invariant masses lower than $500\text{ GeV}$ for the models where we expect a large amount of new heavy particles.

Another interesting application is to search for signatures with missing energy, namely $ZZ + \text{MET}$, $WW + \text{MET}$ or $Z\gamma + \text{MET}$. The first two signatures were considered in the context of SUSY searches at the LHC~\cite{ATLAS:2017uun,Sirunyan:2018ubx}. In our case, unlike the case considered in~\cite{ATLAS:2017uun,Sirunyan:2018ubx} we expect a peak in the number of events corresponding to the invariant mass equal to the mass of the heavy scalar $M_s$. However, this cannot be used to increase sensitivity as only the transverse component of the missing energy can be measured. Alternatively, using the transverse mass of the visible system could provide means to discriminate the considered model from SM backgrounds.

On the other hand, the dedicated search in the channel $Z\gamma + \text{MET}$ was not performed at the LHC. Indeed, in~\cite{Sirunyan:2017nyt,Aaboud:2018doq} the analysis in the channel jets$+\gamma + \text{MET}$ was reported. However, the specification of jets to $Z$ or considering leptonic $Z$ decays should significantly increase sensitivity. As we show in section~\ref{sec:MET}, the signal in this channel is not constrained by $\gamma$+MET search~\cite{Sirunyan:2017ewk,Aaboud:2017dor}, therefore the signal in this channel can be observed. An advantage of this channel as compared to $ZZ + \text{MET}$ or $WW + \text{MET}$ is the high efficiency of reconstruction of high energy photons. The SM background is also expected to be lower. 
We conclude that dedicated searches in the $Z\gamma + \text{MET}$ channel have a potential to discover new physics at the LHC.

\myparagraph{Acknowledgements.} This research was partially supported by the Netherlands Science Foundation (NWO/OCW) and the  European  Research  Council  (grant number 694896).

\appendix

\section{Misidentification of two photons as one photon}
\label{sec:misidentification}

Consider an ultrarelativistic particle with energy $E$ and mass $m$ that decays into 2 photons. Such particle should have spin 0 or 2 (the case of spin 1 is forbidden because of Landau-Yang theorem~\cite{Landau:1948kw,Yang:1950rg}).

The distribution of photons in the rest frame of the decaying particle is isotropic, while in the laboratory frame with the Lorentz factor $\gamma=E/m$, the distribution of the photon pair $N_{\gamma}$ is
\begin{equation}
  \frac{dN_\gamma}{d\theta} = \frac{1}{2\sqrt{\gamma^2 -1}}\frac{\cos(\theta/2)}{\sin^2(\theta/2)}\frac1{\sqrt{\gamma^2 \sin^2(\theta/2) - 1}}\;,
  \label{eq:photon_distribution}
\end{equation}
where $\theta$ is the angle between two photons. The minimal angle between two photons is therefore
\begin{equation}
  \theta_{min} = 2\arcsin(\gamma^{-1}) \approx \frac2\gamma \quad\text{for}\quad \gamma \gg 1\;.
\end{equation}
The distribution~(\ref{eq:photon_distribution}) is sharply peaking and $95\%$ of all events have the angle between the photons $\theta_{min} < \theta < 3\theta_{min}$. Thus, the opening angle most likely is in the region
\begin{equation}
 \theta \lesssim \frac{6 m}{E}.
 \label{eq:deltatheta}
\end{equation}
The mis-identification probability depends on the granularity of the calorimeter used.

\section{Simple UV completion}
\label{sec:toymodel}


Consider the model with $N_f$ heavy fermion doublets $\chi_I$, which are charged by $U_Y(1)$ and $SU_L(2)$ groups of the SM, and the complex field $\phi$ that interacts with them through the Yukawa interaction,
\begin{equation}
    \mathcal{L}_{\chi} = \partial_{\mu}\phi \partial^{\mu}\phi^* - 
    V(\phi)
    + i \bar{\chi}_I\slashed{D}\chi_I 
    -
    m_\chi \bar{\chi}_I\chi_I - 
    \left(y_{IJ}\phi \bar{\chi}_I\chi_J + \text{h.c.}\right),
\end{equation}
where $V(\phi)$ is a scalar potential that makes a spontaneous symmetry breaking and produces heavy scalar $s$ and light pseudogoldstone boson $a$. These states interact with the SM trough the effective coupling~\eqref{eq:lagrangian} made by the fermionic loop. The expected coupling strength depends on the details of the theory but should be of order $c_{1,2}\approx \alpha_w N_f$ for Yakawa values of order one.

\mbox{}

\section{Decay widths of axion and heavy scalar particle}
\label{sec:axion_decay_length}

For the axion $a$ we have the following decay width
\begin{align}
 \Gamma_{a \rightarrow \gamma \gamma} &= 
 \frac{ m_a^3}{16 \pi f^2} \left(c_1 \sin^2 \theta_W + c_2 \cos^2 \theta_W\right)^2. 
 \label{eq:agammagamma} 
\end{align}
In general case without degeneracies,
\begin{equation}
 \frac{c_1 \sin^2 \theta_W + c_2 \cos^2 \theta_W}{f} \sim \frac{|c_1 - c_2|}{f} < 1.6 \cdot 10^{-4}\text{ GeV}^{-1}.
\end{equation}
The estimation for the decay width for such value is
\begin{equation}
 \Gamma_{a \rightarrow \gamma \gamma} = 2\cdot 10^{-13}\text{ GeV} \left( \frac{m_a}{100\text{ MeV}}\right)^3
 \left( \frac{c_1 \sin^2 \theta_W + c_2 \cos^2 \theta_W}{f \cdot 10^{-4} \text{ GeV}^{-1}} \right)^2
\end{equation}

The decay length is given by $l = c\gamma \tau$, where $\tau = \hbar/\Gamma$ is a lifetime, $\gamma$ is a Lorentz factor. Taking the Lorentz factor as $\gamma = M_s / (2 m_a)$ one gets
\begin{equation}
 l = 5\text{ m} \left( \frac{100\text{ MeV}}{m_a}\right)^4
 \left( \frac{M_s}{1 \text{ TeV}}\right)
 \left( \frac{f \cdot 10^{-4} \text{ GeV}^{-1}}{c_1 \sin^2 \theta_W + c_2 \cos^2 \theta_W} \right)^2.
\end{equation}

\bibliographystyle{JHEP} %
\bibliography{3boson} %

\end{document}